\documentclass{ifacconf}

\usepackage{shortex}

\newcommand{\YK}{Youla-Ku\v cera}
\graphicspath{ {figures/} }
\begin{document}

\begin{frontmatter}

\title{A modular framework for stabilizing deep reinforcement learning control} 
\thanks[footnoteinfo]{\textcopyright 2023 the authors. This work has been accepted to IFAC World Congress for publication under a Creative Commons Licence CC-BY-NC-ND.}

% use \thanksref{footnoteinfo} in title
%\thanks[footnoteinfo]{Sponsor and financial support acknowledgment
%goes here. Paper titles should be written in uppercase and lowercase
%letters, not all uppercase.}

\author[math]{Nathan P. Lawrence} 
\author[math]{Philip D. Loewen}
\author[chbe]{Shuyuan Wang}
\author[hw]{Michael G. Forbes}
\author[chbe]{R. Bhushan Gopaluni}
\address[math]{Department of Mathematics, University of British Columbia, Vancouver, BC V6T 1Z2, Canada (e-mail: input@nplawrence.com, loew@math.ubc.ca).}
\address[chbe]{Department of Chemical and Biological Engineering, University of British Columbia, Vancouver, BC V6T 1Z3, Canada (e-mail: antergravity@gmail.com, bhushan.gopaluni@ubc.ca)}
\address[hw]{Honeywell Process Solutions, North Vancouver, BC V7J 3S4, Canada (e-mail: michael.forbes@honeywell.com)}

\begin{abstract}                % Abstract of not more than 250 words.
% Deep reinforcement learning (RL) is an optimization-driven framework for solving general decision making problems.
% Although deep RL has been demonstrated to be an 
%effective tool for feedback control problems, it is also prone to instability. On the other hand, the \YK\ parameterization establishes the set of stabilizing controllers for a given system, effectively disentangling stability and optimization.
%
We propose a framework for the design of feedback controllers that combines the
optimization-driven and model-free advantages of deep reinforcement learning
with the stability guarantees provided by using the \YK\ parameterization to define
the search domain.
Recent advances in behavioral systems allow us to construct a data-driven internal model;
this enables an alternative realization of the \YK\ parameterization based entirely on 
input-output exploration data. 
Using a neural network to express a parameterized set of nonlinear stable operators
enables seamless integration with standard deep learning libraries.
%A key consequence is that stability is not tied to the choice of RL algorithm or its hyperparameters.
We demonstrate the approach on a realistic simulation of a two-tank system.
% 
% 
% 
% 
% We propose a framework that preserves the flexibility and expressive capabilities of general deep RL algorithms, while maintaining the closed-loop stability ``by design'' feature of \YK\ .
% Towards this end, we draw from recent techniques in the behavioral systems literature to construct a data-driven internal model; this enables an alternative realization of the \YK\ parameterization entirely from input-output exploration data. Finally, we construct a parameterized set of nonlinear stable operators using neural networks to enable seamless integration with deep learning libraries.
%A key consequence is that stability is not tied to the choice of RL algorithm or its hyperparameters.
% We demonstrate the approach on a realistic simulation of a two-tank system.
\end{abstract}

\begin{keyword}
Reinforcement learning \sep data-driven control \sep \YK\  parameterization \sep neural networks \sep stability \sep process control
\end{keyword}

\end{frontmatter}
%===============================================================================

\section{Introduction}
\label{sec:intro}

Closed-loop stability is a basic requirement in controller design. 
However, many learning-based control schemes do not address it explicitly \citep{busoniu2018ReinforcementLearning}.
This is somewhat understandable.
First, the ``model-free'' setup assumed in such algorithms, compounded by the complexity of the methods and their
underlying data structures, makes stability difficult to reason about.
Second, especially in the case of reinforcement learning (RL), many of the striking recent success stories pertain
to simulated tasks or game-playing environments in which catastrophic failure has no real-world impact.
% A simple example of these phenomena is understood through 
When the feedback controller is to be learned directly with RL, tuning the discount factor and/or the reward function influences not only the learning performance but also the stability during exploration~\citep{busoniu2018ReinforcementLearning}.
This issue provides a counterpoint to the generality and expressive capacity of modern RL algorithms, which have nonetheless attracted immense interest for control tasks \citep{nian2020ReviewReinforcement}.\par

In this work, we propose a \emph{stability-preserving framework} for RL-based controller design.
Our inspiration is the \YK\  parameterization \citep{anderson1998YoulaKucera}, which gives a characterization of all stabilizing controllers for a given system. 
The key design variable is then a stable ``parameter'', rather than a direct controller representation. 
Optimizing over stable operators is still non-trivial, 
but we show how this can be done in a flexible manner using neural networks. 
Finally, the \YK\ parameterization requires an internal model of the system, which contradicts one of the key advantages of RL.
We address this with tools from the behavioral systems literature \citep{markovsky2021BehavioralSystems}, specifically, Willems' fundamental lemma \citep{willems2005NotePersistency}. 
This powerful result provides a characterization of system dynamics entirely from input-output data. 
We leverage this to yield a ``model-free'' internal representation for the plant, resulting in a mathematically equivalent realization of the \YK\  parameterization.\par

In sum, we disentangle three key components in RL-based control system design: algorithms, function approximators, and dynamic models. 
The resulting framework supports a modular approach to learning stabilizing policies, in which advances in any single category can be applied to improve overall results.

\begin{figure}
\begin{center}
\includegraphics[width=8.4cm]{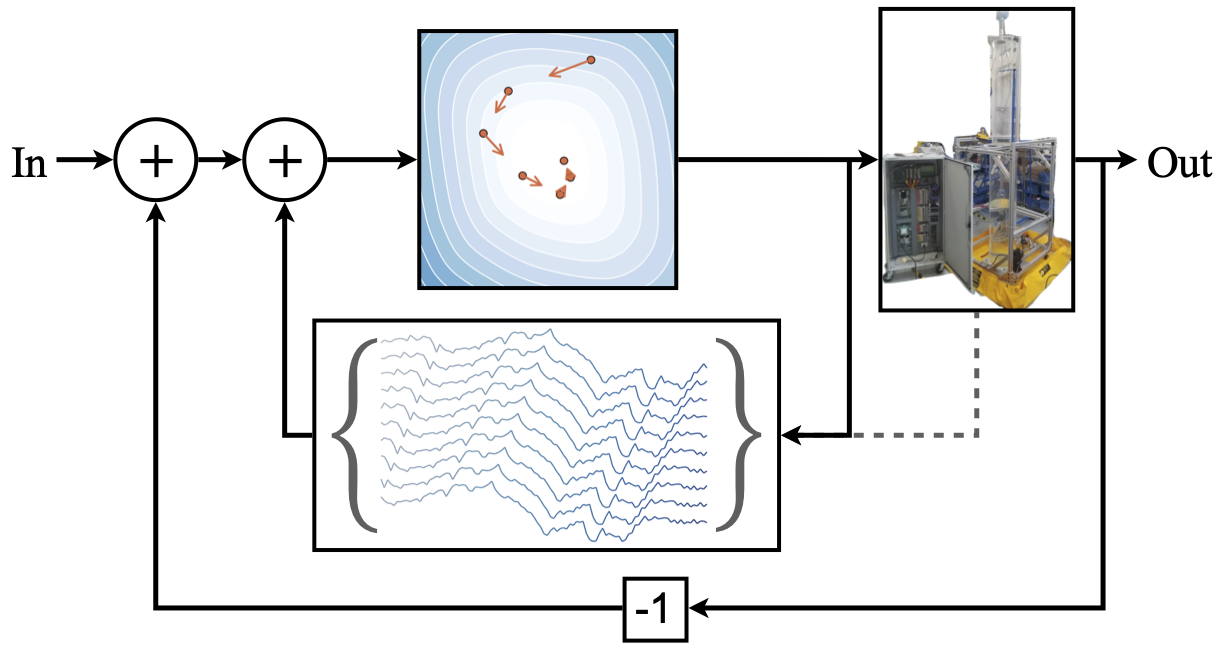}    % The printed column width is 8.4 cm.
\caption{A stable nonlinear parameter $Q$ interacts with its environment;
collected input-output trajectories are used to construct a Hankel matrix.
These ingredients yield an equivalent realization of the \YK\ parameterization.} 
\label{fig:diagram}
\end{center}
\end{figure}

\subsection{Related work}

\citet{busoniu2018ReinforcementLearning} provide a survey of RL techniques from a control-theoretic perspective, 
emphasizing the need for stability-aware RL algorithms. 
Existing strategies for incorporating stability into RL can be described in three broad categories:
integral quadratic constraints (IQCs), Lyapunov's second method, and the \YK\  parameterization.\par

IQCs are a method from robust control theory for proving stability of a dynamical system with some nonlinear or time-varying component. In the context of RL, nonlinearities in the environment or the nonlinear activation functions used to compose a policy neural network can be characterized using IQCs. This has been the approach in several recent works, for example, \citet{jin2020StabilitycertifiedReinforcement,wang2022LearningAll}.\par

Lyapunov stability theory is also well represented in the RL literature \citep{berkenkamp2017SafeModelbased, han2020ActorCriticReinforcement}. The principal idea is to learn a policy that guarantees the steady decrease of a suitable Lyapunov function. \citet{berkenkamp2017SafeModelbased} proposed one of the first methods to establish stability with deep neural network policies: a Lyapunov function and a statistical model of the environment are assumed to be available, then the policy is optimized within an expanding estimate of the region of attraction. Subsequent works add the task of acquiring a Lyapunov function to the learning process \citep{lawrence2020AlmostSurely}. For example, \citet{han2020ActorCriticReinforcement} exploit a trainable Lyapunov neural network in tandem with the policy. Methods based on merging model predictive control with RL \citep{zanon2020SafeReinforcement} also make essential use of Lyapunov analysis.\par
 
The \YK\  parameterization is a seemingly under-utilized technique for integrating stability into RL algorithms.
% \citet{roberts2011FeedbackController} propose its use after evaluating the performance of RL with several different controller parameterizations for a simulated ball-catching task. Subsequently,
\citet{friedrich2017RobustStability} employ the \YK\  parameterization through the use of a crude plant model; RL is used to optimize the tracking performance of a physical two degree of freedom robot in a safe fashion while accounting for unmodeled nonlinearities. Recently, a recurrent neural network architecture based on IQCs was developed \citep{revay2021RecurrentEquilibrium}. Since this architecture satisfies stability conditions by design, it can be used for control in a nonlinear version of the \YK\  parameterization \citep{wang2022LearningAll}.\par

While we also use the \YK\  parameterization, our approach has several novel aspects. Its method for producing stable operators uses a non-recurrent neural network structure; this makes the implementation and integration with off-the-shelf RL algorithms relatively straightforward, for both on-policy and off-policy learning. We also formulate a data-driven realization of the \YK\  parameterization based on Willems' fundamental lemma, essentially removing the prior modeling assumption.

\section{Background}
\label{sec:background}

This section lays out the foundational pieces for our approach. We first connect Willems' lemma to the \YK\ parameterization, then we show our approach to learning stable operators.\par

We consider linear time-invariant (LTI) systems of the form
\begin{align}
\begin{split}
	x_{t+1} &= A x_{t} + B u_{t}\\
	y_{t} &= C x_{t} + D u_{t}
\end{split}
\label{eq:LTI}
\end{align}

Sometimes it is convenient to express \cref{eq:LTI} as a transfer function, in which case we write $P = P(z) = C (z I - A)^{-1} B + D$. We assume that $(A, B)$ is controllable, that $(A,C)$ is observable, and that an upper bound of the order of the system is known. Crucially, the system matrices are unknown. For simplicity in our formulation we assume the system of interest is stable and single-input single-output, however, the results can be extended to more general cases.\par

\subsection{Data-driven realization of the \YK\ parameterization}

Given an $N$-element sequence $\{ z_{t} \}_{t = 0}^{N-1}$ of vectors in $\reals^m$, the \emph{Hankel matrix of order $L$} is given by
\[
H_{L}(z) =
\begin{bmatrix}
	z_{0} & z_{1} & \ldots & z_{N-L} \\
	z_{1} & z_{2} & \ldots & z_{N-L+1} \\
	\vdots & \vdots & \ddots & \vdots \\
	z_{L-1} & z_{L} & \ldots & z_{N-1}
\end{bmatrix}.
\label{eq:Hankel}
\]

% The excitation of a signal can be characterized with the Hankel matrix.

\begin{definition}
The sequence $\{ z_{t} \}_{t = 0}^{N-1} \subset \reals^{m}$ is \emph{persistently exciting of order $L$\/} if $\rank(H_L(z)) = mL$.
\end{definition}

\begin{definition}
An input-output sequence $\{ u_{t}, y_{t} \}_{t = 0}^{N-1}$ is a \emph{trajectory} of an LTI system $(A, B, C, D)$ if there exists a state sequence $\{ x_{t} \}_{t = 0}^{N-1}$ such that \cref{eq:LTI} holds.
\label[definition]{def:trajectory}
\end{definition}

The following theorem is the state-space version of Willems' fundamental lemma \citep{willems2005NotePersistency}. It provides an alternative characterization of an LTI system based entirely on input-output data. Only an upper bound of the order of the system is required. 

\begin{theorem}[See \citet{vanwaarde2020WillemsFundamental}]
Let $\{ u_{t}, y_{t} \}_{t = 0}^{N-1}$ be a trajectory of an LTI system $(A, B, C, D)$ where $u$ is persistently exciting of order $L+n$. Then $\{ \bar{u}_{t}, \bar{y}_{t} \}_{t = 0}^{L-1}$ is a trajectory of $(A, B, C, D)$ if and only if there exists $\alpha \in \reals^{N-L+1}$ such that
\begin{equation}
\begin{bmatrix}
	H_L(u) \\
	H_L(y) 
\end{bmatrix}
\alpha =
\begin{bmatrix}
\bar{u} \\
\bar{y}	
\end{bmatrix}.
\label{eq:fundamentalLemma}
\end{equation}
\label{thm:fundamentalLemma}
\end{theorem}

(When we omit the time index in the context of the right-hand side of \cref{eq:fundamentalLemma}, it is understood as a column vector $\bar{z} = [\bar{z}_0 \ldots \bar{z}_{L-1}]^T$. )
\Cref{thm:fundamentalLemma} has been applied extensively for predictive control tasks \citep{markovsky2021BehavioralSystems,berberich2020TrajectorybasedFramework}. In particular, a sequence of inputs may be proposed, and through a slight variation of \cref{eq:fundamentalLemma}, the corresponding outputs computed. This leads to a scheme of forecasting a sequence of inputs and ``filling in'' the outputs.\par
In what follows, we consider the scenario of using the Hankel based model as an internal system model. Therefore, we assume the system is strictly proper --- that is, $D = 0$ in \cref{eq:LTI} --- to ensure a realizable controller strategy later on. Now, given a system trajectory $\{ u_{t}, y_{t} \}_{t = 0}^{L-1}$, we note that $y_{L}$ is uniquely determined by these available data. A simple way of stepping the system forward is to consider a time-shifted Hankel matrix
\[
H'_{L}(z) = H_{L}(z'),
\]
where $z = \{ z_{t} \}_{t = 0}^{N-1}$ and $z' = \{ z_{t} \}_{t = 1}^{N}$ for some $N$.

\begin{corollary}\label[corollary]{cor:proper}
Let $\{ u_{t}, y_{t} \}_{t = 0}^{N-1}$ be a trajectory of a strictly proper LTI system $(A, B, C)$ where $u$ is persistently exciting of order $L+n+1$. Then for each trajectory $\{ \bar{u}_{t}, \bar{y}_{t} \}_{t = 0}^{L-1}$ of $(A, B, C)$, there exists $\alpha \in \reals^{N-L}$ such that
%\[
%\begin{bmatrix}
%	\bar{y}\\
%	\bar{y}_{L}
%\end{bmatrix}
%= H_{L+1}(y)\alpha. 
%\label{eq:proper}
%\]
\[
\bar{y}'
= H'_{L}(y)\alpha. 
\label{eq:proper}
\]
\end{corollary}

\begin{pf}
By \cref{thm:fundamentalLemma}, the trajectory $\{ \bar{u}_{t}, \bar{y}_{t} \}_{t = 0}^{L-1}$ satisfies
\[
\begin{bmatrix}
	H_L(u) \\
	H_L(y) 
\end{bmatrix}
\alpha =
\begin{bmatrix}
\bar{u} \\
\bar{y}	
\end{bmatrix}
\]
for some $\alpha \in \reals^{N-L}$. Moreover, by \cref{def:trajectory} there exists a sequence of states $\{\bar{x}_{t}\}_{t=0}^{L-1}$ that corresponds to the input-output trajectory $\{ \bar{u}_{t}, \bar{y}_{t} \}_{t = 0}^{L-1}$. This sequence induces the state $\bar{x}_{L}$. We have
\begin{align*}
	\bar{y}_{L} &= C \bar{x}_{L} \\
	&= C\left( A \bar{x}_{L-1} + B \bar{u}_{L-1} \right) \\
	&= C \left( A \sum_{i=0}^{N-L-1} \alpha_{i} x_{L+i} + B \sum_{i=0}^{N-L-1} \alpha_{i} u_{L+i} \right) \\
	&= \sum_{i=0}^{N-L-1} \alpha_i C \left( A x_{L+i} + B u_{L+i} \right) \\
	& = \sum_{i=0}^{N-L-1} \alpha_i y_{L+i+1} \\
\end{align*}
as desired. \qed
\end{pf}

\cref{cor:proper} gives a systematic way of stepping a trajectory forward in time. This is particularly useful for aligning the true system with a Hankel representation while implementing a feedback controller online.\par

The \YK\  parameterization produces the set of all stabilizing controllers through a combination of an internal system model and a stable operator. The trick is to directly parameterize the closed-loop transfer functions associated with the plant, then recover a controller. For example, the behavior of the closed-loop transfer function $\frac{P C}{1 + P C}$ from the reference $r$ to output $y$ is determined by the transfer function $\frac{C}{1 + P C}$. By introducing a stable design variable $Q$, we can then directly shape the stable behavior of the system through the transfer function $PQ$. By setting $Q = \frac{C}{1 + P C}$, we arrive at the \YK\  parameterization \citep{anderson1998YoulaKucera}:
\[
\mathcal{C}_{\text{stable}} = \left\{ \frac{Q}{1 - Q P} \colon Q \text{ is stable} \right\}
\label{eq:YK}
\]
This result extends further to unstable, multiple-input multiple-output systems. Moreover, when $P$ is linear, one may use a nonlinear operator $Q$ \citep{anderson1998YoulaKucera}. \par

%{
%\begin{minipage}[]{\linewidth}
%\linespread{1}
%\normalsize
%\SetArgSty{textnormal}
%\begin{algorithm}[H]
%\caption{Data-driven stabilizing controller}\label{alg:YK}
%\SetKwInput{KwInput}{Input}
%\KwInput{Stable $Q$ parameter; Parameter $L+1$; Data $\{u_k, y_k\}_{k=0}^{N-1}$; Initial trajectory $\{\bar{u}_k, \bar{y}_k\}_{k = 0}^{L-1}$}  
%\For{each time step $t$}{
%	Set $u_{t-1} \gets \bar{u}_{L-1}$\;
%	Observe the tracking error $e_t = r_t - y_{t}$ from the system\;
%	Compute $\bar{y}_L$ from \cref{eq:proper}\;
%	Apply the input $\hat{r} = e_t + \bar{y}_{L}$ to the $Q$ parameter and return control action $u_{L}$\;
%	Update trajectory: \;
%	 \nonl$$
%	 	\{\bar{u}_k, \bar{y}_k\}_{k = 0}^{L-1} \gets \{\bar{u}_k, \bar{y}_k\}_{k = 1}^{L} $$\;	
%}
%\end{algorithm}
%\end{minipage}
%}

In \cref{alg:YK}, we translate the mathematical ideas above into a direct
sequential process.
The following result provides details of the correspondence.

%Assume $P$ is a stable and strictly proper \ac{LTI} system. Let $Q$ be a stable and proper \ac{LTI} parameter. Given an upper bound $L$ of the order of $P$, \cref{alg:YK} produces the same control signal $\{\bar{u}_t\}_{t=0}^{\infty}$ as the \YK\  parameterization.
\begin{theorem}
Assume $P$ is a stable and strictly proper LTI system. Let $Q$ be a stable and proper LTI parameter. Given an upper bound $L$ of the order of $P$, \cref{alg:YK} produces the same control signal $\{\bar{u}_t\}_{t=0}^{\infty}$ as the \YK\  parameterization.
\end{theorem}
\begin{pf}
%The \YK\  parameterization tells us that the controller $C = \frac{U}{E} = \frac{Q}{1 - PQ}$ is stabilizing. Equivalently,
%\begin{align}
%	U &= PQU + QE \\
%	\iff u_t &= p(t)*(q*u)(t) + (q*e)(t) \\
%\end{align}
We use $q_t$, $p_t$ to denote the impulse responses of $Q$ and $P$, respectively. Similarly, respective minimal state-space matrices are denoted $(A_{q}, B_{q}, C_{q}, D_{q})$ and $(A_{p}, B_{p}, C_{p})$.\par 
By the \YK\  parameterization, we have
\begin{align}
&& C(z) &= \frac{Q(z)}{1 - Q(z)P(z)}\quad \forall z \in \comps\nonumber \\
\iff&& \left(1 - Q(z)P(z) \right) U(z) &= Q(z) E(z)\nonumber \\
\iff&& u_t &= q_t * (e_t + p_t * u_t)\quad  \forall t \in \natsO\nonumber \\
&& &= \sum_{j=0}^{t-1} C_{q} A_{q}^{t-1-j} B_{q} \hat{r}_{j} + D_{q} \hat{r}_{t}, \label{eq:YKimpulse}
\end{align}
where $\hat{r}_{j} = e_{j} + \sum_{i=0}^{j-1} C_{p}A_{p}^{j-1-i}B_{p} u_i$ and $*$ is the convolution operator; we have also assumed, without loss of generality, that $P$ and $Q$ have zero initial state.\par
Next we relate \cref{eq:YKimpulse} to \cref{alg:YK}.
Let $\{ e_k \}_{k=0}^{\infty}$ be an arbitrary sequence. (Such a sequence is dynamically generated in \cref{alg:YK}.)
Without loss of generality, let the initial trajectory be $\{\bar{u}_k, \bar{y}_k\}_{k = 0}^{L-1} = \{0, 0\}_{k = 0}^{L-1}$. For each time $t \in \natsO$ we compute $\alpha^{(t)}$ and $\bar{y}_{t} = \bar{y}_{L}$ from \cref{eq:proper}. Since $L$ is an upper bound of the order of $P$, $\bar{y}_{t}$ is the unique next output from the trajectory $\{\bar{u}_k, \bar{y}_k\}_{k = 0}^{L-1}$. Therefore, we have 
\[
	\hat{r}_{t} = e_{t} + \sum_{i=0}^{N-L-1} \alpha_{i}^{(t)} y_{L+i+1}.
\]
Then
\[
	\bar{u}_t = \sum_{j=0}^{t-1} C_{q} A_{q}^{t-1-j} B_{q} \hat{r}_{j} + D_{q} \hat{r}_{t}
\]
gives the next control input. \par
By updating the trajectory between time steps --- $\{\bar{u}_k, \bar{y}_k\}_{k = 0}^{L-1} \gets \{\bar{u}_k, \bar{y}_k\}_{k = 1}^{L}$ --- we dynamically generate a sequence $\{ \alpha^{(t)} \}_{t=0}^{\infty}$ that produces the control inputs $\{ \bar{u}_{t} \}_{t=0}^{\infty}$ satisfying the discrete integral equation in \cref{eq:YKimpulse}. \qed
\end{pf}

\begin{algorithm}[t!]
\caption{Data-driven stabilizing controller}\label{alg:YK}
\begin{algorithmic}[1]
	\State \textbf{Input:} Stable $Q$ parameter; Observations $\{u_k, y_k\}_{k=0}^{N-1}$; Initial trajectory $\{\bar{u}_k, \bar{y}_k\}_{k = 0}^{L-1}$
	\For{{each time step $t$}}{}{}
		\State Set $u_{t-1} \gets \bar{u}_{L-1}$
		\State Observe the tracking error $e_t = r_t - y_{t}$
		\State Compute $\bar{y}_L$ from \cref{eq:proper}
		\State Apply the input $\hat{r} = e_t + \bar{y}_{L}$ to the $Q$ parameter and return control action $u_{L}$
		\State Update trajectory: $$\{\bar{u}_k, \bar{y}_k\}_{k = 0}^{L-1} \gets \{\bar{u}_k, \bar{y}_k\}_{k = 1}^{L} $$
	\EndFor
\end{algorithmic}
\end{algorithm}

\subsection{Learning stable operators}
\label{subsec:Qparam}

The \YK\  parameterization is very elegant, as it refines the search space for any problem to the set of stable operators. However, effectively optimizing over this set is still a major challenge \citep{wang2022LearningAll}.\par

We adapt the method due to \citet{lawrence2020AlmostSurely}. First let us recall the notion of a \emph{Lyapunov candidate function} $V\colon \reals^n \to \reals$: 1) $V$ is continuous; 2) $V(z) > 0$ for all $z \neq 0$, and $V(0) = 0$; 3) There exists a continuous, strictly increasing function $\varphi\colon[0, \infty) \to [0, \infty)$ such that $V(z) \geq \varphi(\norm{z})$ for all $z \in \reals^n$; 4) $V(z) \to \infty$ as $\norm{z} \to \infty$.\par 
%\begin{enumerate}
%    \item $V\colon\reals^n \to \reals$ is continuous
%    \item $V(x) > 0$ for all $x \neq 0$, and $V(0) = 0$
%    \item There exists a continuous, strictly increasing function $\varphi\colon[0, \infty) \to [0, \infty)$ such that $V(x) \geq \varphi(\norm{x})$ for all $x \in \reals^n$
%    \item $V(x) \to \infty$ as $\norm{x} \to \infty$
%\end{enumerate}
Lyapunov functions are instrumental for proving a system is stable through Lyapunov's second method \citep{khalil2002nonlinear}. \citet{lawrence2020AlmostSurely} construct stable autonomous systems of the form $z_{t+1} = f_{\theta}(z_t)$ ``by design'' through the use of trainable Lyapunov functions. A neural network satisfying the principal requirements above can be obtained through a slightly modified input-convex neural network \citep{amos2017InputConvex} --- see \citet{lawrence2020AlmostSurely} and the references therein for details. \par

%For the moment we focus on learning stable autonomous systems of the form $x_{t+1} = f_{\theta}(x_t)$. The main idea is to construct a stable trainable model ``by design''. The method uses two neural networks: a smooth neural network $\hat{f}_{\theta}$, and a Lyapunov neural network $V_{\theta}$. A neural network satisfying the conditions of a Lyapunov candidate function can be constructed with the use of an input convex neural network \citep{amos2017InputConvexa}.

Two neural networks work in tandem to form a single model that satisfies the decrease condition central to Lyapunov's second method: a smooth neural network $\hat{f}_{\theta}$, and a Lyapunov neural network $V_{\theta}$. Set $\hat{z}' = \hat{f}_{\theta}(z)$ where $z$ is the current state and $\hat{z}'$ is the proposed next state. Two cases are possible: either $\hat{z}'$ decreased the value of $V$ or it did not. We can write out a ``correction'' to the dynamics in closed form by exploiting the convexity of $V$:
\begin{align}
\begin{split}
    z_{t+1} &= f_{\theta}( z_t )\\
    & \equiv
    \begin{cases}
    \hat{f}_{\theta}(z_t), &\text{if } V( \hat{f}_{\theta}(z_t) ) \leq \beta V( z_t )\\
    \hat{f}_{\theta}(z_t)\left( \frac{\beta V( z_t )}{V( \hat{f}_{\theta}(z_t) )} \right), &\text{otherwise}
    \end{cases}\\
     &= \gamma \hat{f}_{\theta}( z_t ), \text{ where}\\\
     & \gamma=\gamma( z_t )= \frac{\beta V( z_t ) - \texttt{ReLU}\big( \beta V( z_t ) - V( \hat{f}_{\theta}( z_t ) )}{ V( \hat{f}_{\theta}( z_t ))}.
\end{split}
\label{eq:det_stable_scale}
\end{align}

Since \cref{eq:det_stable_scale} defines the model $f_{\theta}$, both $\hat{f}_{\theta}$ and $V_{\theta}$ are trained in unison towards whatever goal is required of the sequential states $z_t, z_{t+1}, \ldots$, such as supervised learning tasks. Moreover, although the model $f_{\theta}$ is constrained to be stable, it is unconstrained in parameter space, making its implementation and training fairly straightforward with deep learning libraries. Further, despite the complex structure in \cref{eq:det_stable_scale}, \citet{lawrence2020AlmostSurely} show that the overall model is continuous. Here, we use $f_{\theta}$ to model the internal dynamics of a nonlinear $Q$ parameter. For example, a control--affine model may be used with stable transition dynamics $f_{\theta}$.

\section{Unconstrained stabilizing reinforcement learning}

A brief overview of deep RL will serve to define our notation, which is largely standard.
For more background, see \citet{sutton2018ReinforcementLearning, busoniu2018ReinforcementLearning}; a tutorial-style treatment is given by \citet{nian2020ReviewReinforcement}.
\par

Reinforcement learning is an optimization-driven framework for learning ``policies'' simply through interactions with an environment \citep{sutton2018ReinforcementLearning}. The states $s$ and actions $a$ belong to the state and action sets $\mathcal{S}$, $\mathcal{A}$, respectively. At each time step $t$, the state $s_t$ influences the sampling of an action $a_t \sim \pi(\cdot \mid s_t)$ from the ``policy'' $\pi$. Given the action $a_t$, the environment produces a successor state $s_{t+1}$. This cycle completes one step in a Markov decision process, which induces a conditional density function $s_{t+1} \sim p( \cdot \mid s_t, a_t )$ for any initial distribution $s_{0} \sim p_{0}( \cdot )$. As time steps forward under a policy $\pi$, a ``rollout'' is denoted $h = (s_0, a_0, r_0, s_1, a_1, r_1, \ldots )$. Each fixed policy $\pi$, induces a probability density $p^\pi(\cdot)$ on the set of trajectories.\par

In RL the desirability of a given rollout is quantified by a ``reward'' $r_t = r(s_t, a_t)$ associated
with each stage in the process above.
The overall goal of the agent is to determine a policy that maximizes the cumulative discounted reward.
That is, given some constant $\gamma\in(0,1)$, 
\begin{equation}
\begin{aligned}
    &\text{maximize} && J(\pi) = \mathbb{E}_{h \sim p^{\pi}}\left[ \sum_{t=0}^{\infty} \gamma^{t}r(s_t,a_t) \right]\\
    &\text{over all} && \text{policies } \pi \colon \mathcal{S} \to \mathcal{P}(\mathcal{A}),
\end{aligned}
\label{eq:RLobjective}
\end{equation}
where $\mathcal{P}(\mathcal{A})$ denotes the set of probability measures on $\mathcal{A}$.\par
% In this problem, $h\sim p^\pi$ refers to a typical trajectory
% $h = (s_0, a_0, r_0, , \ldots, s_N, a_N, r_N)$ generated by
% the policy $\pi$ with sub-sequential states distributed according to $p$.

In the space of all possible policies, the optimization is performed over a subset parameterized by some vector $\theta$. For example, in some applications, $\theta$ denotes the set of all weights in a deep neural network. In this work, the policy is the nonlinear $Q$ parameter outlined in \cref{subsec:Qparam}. Therefore, \cref{eq:RLobjective} automatically satisfies an internal stability constraint over the whole weight space $\theta$. We are then able to use any RL algorithm to solve the problem. We do not recount the inner workings of common RL algorithms, as they are well-documented \citep{nian2020ReviewReinforcement}. Instead, a brief overview is given. \par

The broad subject of reinforcement learning concerns iterative methods for choosing a desirable policy $\pi$ (this is the ``learning''), guided in some fundamental way by the agent's observations of the rewards from past state-action pairs (this provides the ``reinforcement'').\par

A standard approach to solving Problem~\eqref{eq:RLobjective} uses gradient ascent
\[
\theta
\leftarrow \theta + \alpha\nabla J(\theta),
\label{eq:PolicyGradient_Iteration}
\]
where $\alpha > 0$ is a step-size parameter.
Analytic expressions for $\nabla J(\theta)$ exist for both stochastic and deterministic policies (see \citet{sutton2018ReinforcementLearning,busoniu2018ReinforcementLearning}). Crucially, these formulas rely on the state-action value function$\!$\footnote{Unfortunately both this function and the \YK\ parameter are typically denoted by $Q$. This explains the superscript here.}$\!\!$,  % or $Q$-function:
\[
    Q^{(\text{RL})}(s_t, a_t) = \mathbb{E}_{h \sim p^\pi}\left[ \sum_{k = t}^{\infty} \gamma^{k-t}r(s_k,a_k) \middle| s_t, a_t \right].
\label{eq:Qfunc}
\]
Although $Q^{(\text{RL})}$ is not known precisely, as it depends on both the dynamics and the policy, it can be estimated with a deep neural network \citep{busoniu2018ReinforcementLearning}.
These ideas and various approximation techniques form the basis of deep RL algorithms.

\section{A simulation example}
\label{sec:example}

\begin{figure}
\begin{center}
\includegraphics[width=8.4cm]{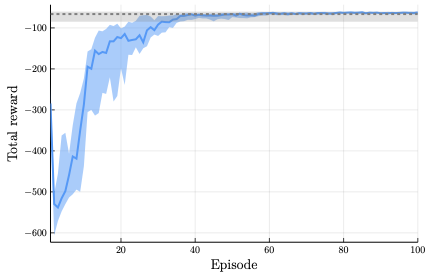}    % The printed column width is 8.4 cm.
\caption{Cumulative reward curve over $20$ training sessions. The solid line is the median and the shaded region shows the interquartile range. The dashed line and its shaded region are the final results of training without the stability constraint.} 
\label{fig:rewards}
\end{center}
\end{figure}

We showcase our training results on a realistic simulation of a level-control system involving
two large tanks of water. 
The objective is to regulate the water level in the upper tank, while water continuously drains out into a lower reservoir. A pump lifts water from the reservoir back to the upper tank, establishing a cyclic flow. A physical depiction of the setup is shown in \cref{fig:diagram} and further explained in \citet{lawrence2022DeepReinforcement}. \par 
The system dynamics are based on Bernoulli's equation, establishing outflow $f_{\text{out}}\approx f_c\sqrt{2 g\ell}$, and the conservation of fluid volume in the upper tank:
\[
\frac{d\hfil}{dt}\left(\pi r_{\text{tank}}^2\ell\right)
= \pi r_{\text{tank}}^{2} \dot\ell
= f_{\text{in}} - f_{\text{out}}
\]
(We use dot notation to represent differentiation with respect to time; $g$ is the gravitational constant; $\ell$ is the level; $r_{\text{tank}}$ is the radius of the tank; see \citet{lawrence2022DeepReinforcement} for a more thorough description of the system.)
% Our system model combines the principles above with some simple filters to make our mathematical description physically realizable. A first-order filter with time constant $\tau\ge0$
% transforms an input signal $\hat{y}$
% into the output signal $y$ defined by
% \begin{equation}
% \tau \dot{y} + y = \hat{y},
% \qquad y(0)=0.
% \end{equation}
% Note that setting $\tau=0$ gives $y(t)=\hat y(t)$, while if $\hat y$ is constant,
% one has $y(t) = (1 - e^{-t/\tau})\hat y$.
Our application involves four filtered signals, with time constants
$\tau_p$ for the pump,
$\tau_{\text{in}}$ for changes in the inflow,
$\tau_{\text{out}}$ for the outflow,  and
$\tau_m$ for the measured level dynamics. We therefore have the following system of differential equations describing the pump speed, flow rates, level, and measured level, respectively:
\begin{align*}
\tau_p \dot{p} + p &= p_{\text{sp}}\\
\tau_{\text{in}} \dot{f}_{\text{in}} + f_{\text{in}} &= f_{\text{max}} \left(\frac{p}{100}\right)\\
\tau_{\text{out}} \dot{f}_{\text{out}}+ f_{\text{out}} &= \pi r_{\text{pipe}}^{2} f_{\text{c}} \sqrt{2 g \ell}\\
\pi r_{\text{tank}}^{2} \dot{\ell} &= f_{\text{in}} -  f_{\text{out}}\\
\tau_m \dot{m} + m &= \ell
\end{align*}
To track a desired level $\ell_{\text{sp}}$ --- ``sp'' stands for ``setpoint'' --- we can employ level and flow controllers by including the following equations:
\begin{align}
\begin{split}
p_{\text{sp}} &= \text{PID}_{\text{flow}} (f_{\text{in,sp}} - f_{\text{in}}) \\
f_{\text{in,sp}} &= \text{PID}_{\text{level}}(\ell_{\text{sp}} - m)
\end{split}
\label{eq:PIDtank}
\end{align}
\Cref{eq:PIDtank} uses shorthand for PID controllers taking the error signals $f_{\text{in,sp}} - f_{\text{in}}$ and $\ell_{\text{sp}} - m$, respectively. For our purposes, $\text{PID}_{\text{flow}}$ and $\text{PID}_{\text{level}}$ are fixed and a part of the environment. The implementation of these dynamics is performed in discrete time steps of 0.5 seconds and with Gaussian measurement noise with variance $0.015$.\par

\begin{figure}
\begin{center}
\includegraphics[width=8.4cm]{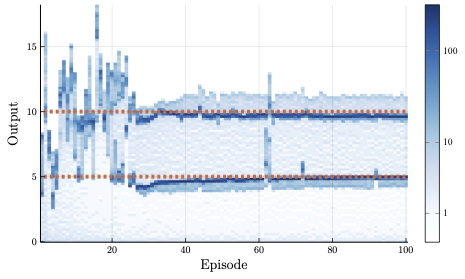}    % The printed column width is 8.4 cm.
\caption{A global view of the training progress across all $20$ sessions. For each episode, a distribution of time spent at various output values is obtained. The heatmap shows the average amount of time spent at each episode--output coordinate.} 
\label{fig:heatmap_output}
\end{center}
\end{figure}

\emph{(Training results)}\quad
Since the environment includes a PID controller, we modify the control scheme to be in incremental form $u_t = u_{t-1} + \Delta u_t$, where $\Delta u_t$ is the sum of the \YK\  parameter and PID controller outputs:
\[
\Delta u_t = \Delta u^{(q)}_t + \Delta u^{(\text{PID})}_t
\]
We ran $20$ training sessions for $100$ episodes each and combined the results in \cref{fig:rewards,fig:heatmap_output}. \Cref{fig:rewards} shows the cumulative rewards for each episode. We convey the median and interquartile ranges over the $20$ training sessions; we see that median reward curve is much closer to the upper limit of the shaded region than the lower, indicating that the majority of experiments fall within that tight region. Although there is significant variation at initialization, due to the random policy initialization, the training sessions exhibit consistent convergence. The reward curves tend to plateau after around $40$ episodes.\par

\begin{figure}
\begin{center}
\includegraphics[width=8.4cm]{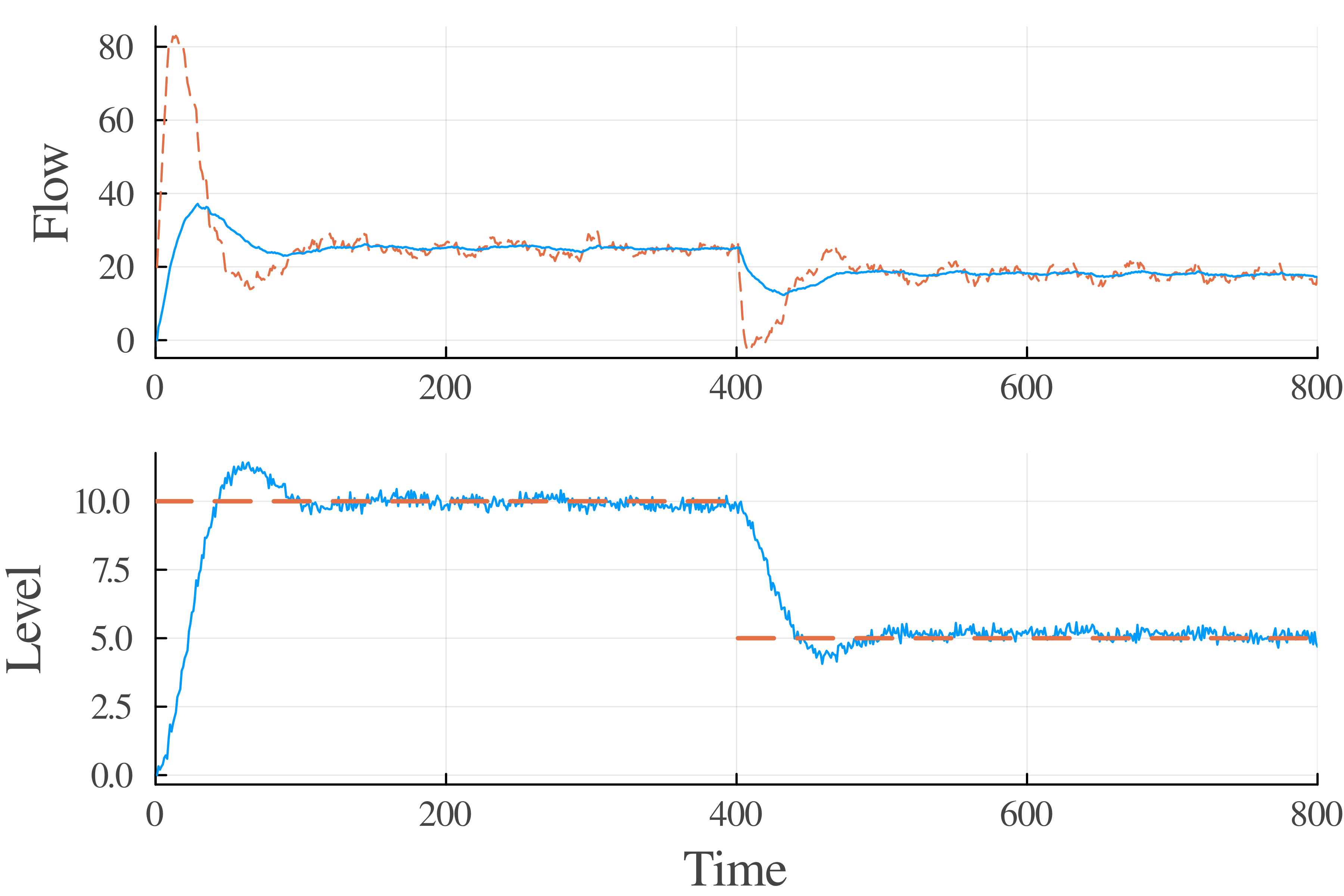}    % The printed column width is 8.4 cm.
\caption{A sample input-output rollout by the trained RL agent for one of the training sessions. Dashed lines are setpoints; solid lines are measured values.} 
\label{fig:input_output}
\end{center}
\end{figure}

\Cref{fig:heatmap_output} shows the collective evolution of each episode throughout the training sessions. Each episode--output coordinate is shaded based on how much time the output variable spent there on average. Darker shading around the dashed values (setpoints) is more desirable. The purpose of this figure is to provide a rough translation of what the reward curve in \cref{fig:rewards} entails. On the other hand, \cref{fig:input_output} shows a single rollout from one of the experiments.

\section{Conclusion}

The \YK\  parameterization is well-known in control theory, but seemingly under-utilized in RL. Taking it as a starting point, we have adopted advances in deep learning and behavioral systems to develop an end-to-end framework for learning stabilizing policies with general RL algorithms. This paper is a proof of concept and there are many avenues to explore. These include the use of stochastic policies; extensions to unstable systems; and balancing the persistence of excitation assumption during training and steady-state operations. We believe this is a fruitful area to investigate further as deep RL gains traction in process systems engineering.

\begin{ack}
We gratefully acknowledge the financial support of the Natural Sciences and Engineering Research Council of Canada (NSERC) and Honeywell Connected Plant. We would also like to thank Professor Yaniv Plan for helpful discussions.
\end{ack}

\bibliography{bibliography}  

\appendix
\section{Implementation details}

Numerical experiments were carried out in the Julia programming language. In particular, we utilized the \texttt{ReinforcementLearning.jl} package \citep{Tian2020Reinforcement}. We used the TD3 algorithm \citep{fujimoto2018AddressingFunction} to update network parameters. We used the reward function $- 0.1\abs{y_{\text{sp}} - y} - 0.01 \left(\Delta u^{(q)} \right)^2$. Most hyperparameters were set to their default values; we set the policy delay to $4$. We used two-layer feedforward networks throughout. The critic network had $64$ nodes per layer and used the \texttt{softplus} activation. We used the same structure for the actor in the stability-free comparison shown in \cref{fig:rewards}. For the stable $Q$ parameter we used \cref{eq:det_stable_scale} and created a state-space model with matrices $B,C,D$ and $f_\theta$ instead of the nominal $A$ matrix. $\hat{f}_{\theta}$ had $16$ nodes per layer and used the \texttt{tanh} activation. $V_\theta$ also had $16$ nodes per layer. 

%\begin{table}[H]
%\caption{Hyperparameters used in our example.}\label{table:hyperparameters}
%\begin{center}
% \begin{tabularx}{\linewidth}{XX}
%\toprule
%    Hyperparameter & Value \\
%\midrule
%Algorithm & TD3 \citep{fujimoto2018AddressingFunction} \\
%Reward & $- 0.1\abs{y_{\text{sp}} - y} - 0.01 \left(\Delta u^{(q)} \right)^2$ \\
%Discount factor & $0.99$ \\
%Target network filter & $0.99$ \\
%Optimizer &  ADAM \\
%Learning rate &  $0.001$ \\
%Batch size & $64$ \\
%Policy delay & $4$ \\
%Replay buffer size & $10,000$ \\
%\bottomrule
%\end{tabularx}
%\end{center}
%\end{table}

%
%\begin{table}[H]
%\caption{Parameters and variables for the two-tank system. Length is in decimeters (dm), time is in minutes, and volume is in liters. The tank height in our simulation is 12.192 dm.}\label{table:tank}
%\begin{tabularx}{\linewidth}{lXX}
%\toprule
%    Symbol &    Value or unit &    Description\\
%\midrule
% $r_{\text{tank}}$ & 1.2065 (length) & Tank radius \\ 
% $r_{\text{pipe}}$ & 0.125 (length) & Outflow pipe radius \\ 
% $f_\text{max}$ & 80 (volume/time) & Maximum flow \\ 
% $f_c$ & 0.61 & Flow coefficient \\ 
% $\tau_{p, \text{in}, \text{out}, m}$ & 0.1, 0.1, 0.1, 0.2 (time) & Time constants \\ 
% % $g$ & (length/$\text{time}^2$) & Gravitational constant \\ 
% % $\ell$ & length & Tank level  \\ 
% % $m$ & length & Filtered tank level \\ 
% % $f_\text{in}$ & volume/time & Inflow  \\ 
% % $f_\text{out}$ & volume/time & Outflow  \\ 
% % $p$ & \% & Pump speed  \\ 
% % $p_{\text{sp}}$ & \% & Desired pump speed  \\ 
%\bottomrule
%\end{tabularx}
%\end{table}

\end{document}